\documentclass[12pt]{iopart}

\usepackage{iopams}  
\usepackage{color, graphicx, amssymb, mathrsfs}  

\begin{document}  
\title{Design concepts for an improved integrated scanning SQUID }

\author{Nicholas C. Koshnick}   
\address{Center for Probing the Nanoscale, Stanford University, Stanford, CA 94305}  
\ead{koshnick@alum.dartmouth.org}  
\author{John R. Kirtley}
\address{Center for Probing the Nanoscale, Stanford University, Stanford, CA 94305}  
\ead{kirtley@ucsbalum.net}
\author{Kathryn A. Moler} 
\address{Center for Probing the Nanoscale, Stanford University, Stanford, CA 94305}  
\ead{kmoler@stanford.edu}  

\begin{abstract} 
In this paper we discuss design concepts for increasing the spatial resolution, improving the sensitivity, and reducing the invasiveness in scanning Superconducting Quantum Interference Device (SQUID) microscope sensors with integrated flux pickup loops. This can be done not only by reducing the ground-rule line widths and spacings, but also by taking advantage of planarization, reducing flux noise through reducing the SQUID inductance, and reducing back-action through dispersive readouts or on-chip filtering. 
\end{abstract}


\maketitle

\section{Introduction}

%
Scanning SQUID microscope sensors with integrated flux pickup loops \cite{ketchenoptimization,kirtley_high-resolution_1995} can have several advantages over sensors composed solely of very small SQUIDs:  they can have reduced interaction with the sample because the junctions can be physically distant from the sample, it is possible to make very small, well shielded pickup loops without having to make the entire device small, and it is comparatively easy to flux modulate them. However, integrated SQUID sensors have the disadvantage of higher complexity, since they require crossovers and overlays for good shielding of the leads to the pickup loop. Integrated SQUID sensors have been very successful in, for example, elucidating the pairing symmetry in the cuprate high temperature superconductors \cite{tsuei_pairing_1994,kirtley2006}, testing the interplane tunneling model for the mechanism of high temperature superconductivity \cite{moler_images_1998}, and imaging spontaneous circulating currents in mesoscopic normal metal rings \cite{bluhm_persistent_2009}. Further, integrated SQUID sensors can be designed in a susceptometer configuration, with a field coil surrounding the pickup loop \cite{gardner_scanning_2001}. This allows for strong discrimination against background fields \cite{bluhm_persistent_2009,koshnick_fluctuation_2007}, manipulation of local magnetic features \cite{gardner_scanning_2001,hilgenkamp_ordering_2003}, and such measurements as imaging of the local superfluid density of a superconductor \cite{tafuri_magnetic_2004}. It is important to minimize the size of the pickup loop and the spacing between the loop and the sample, while maintaining good shielding of the leads to the pickup loop, to optimize sensor spatial resolution and sensitivity.

In previous work \cite{huber_gradiometric_2008, koshnick_terraced_2008}, we presented SQUID susceptometers that were designed explicitly for the purpose of measuring mesoscopic objects.
The first paper in this series \cite{huber_gradiometric_2008} focused on the design's symmetry and on issues relating to operation and performance.  The primary achievement of this generation of devices is the ability to measure the magnetic signals from micron-sized objects down to the fundamental Johnson noise limits \cite{bluhm_persistent_2009}.
The second paper in this series \cite{koshnick_terraced_2008} focussed on reducing the pickup loop size and line width to a point close to the limits set by the superconducting penetration depth of Niobium, the material of choice for superconducting integrated circuits.  This paper also demonstrated a design that avoids the thickness problems associated with the multiple layers required by a shielded device with a pickup loop inside a larger field coil.  Specifically, layer crossings occurred in such a way that the pickup loop could be positioned as close as possible to the surface (Fig \ref{fig:ibmSQ_layers}a).  

In this work, we discuss additional ways to improve SQUIDs for the  purpose of measuring mesoscopic objects as well as imaging.  This revision was motivated, in part, by the possibility of a planarized superconducting process \cite{ketchen_sub-mu_1991}, with design rules that allow for 250 nm optically defined features, and 1 $\mu$m vias.   Section \ref{sec:planar_design} discusses design improvements that would be made possible if this process is implemented.  Planarization allows for significantly reduced geometric constraints in the tip design, and continuous conducting layers that reduce vortex motion, thereby allowing for higher locally applied fields.  
The small feature size allows for a significant reduction in the primary imaging area and in the device's stray pickup area. 
Section \ref{sec:flux_optimization} outlines steps that would reduce the SQUID's inductance below that of previous designs.  This in turn should allow for a reduction in the SQUIDs flux noise.   The most significant reduction in inductance comes from a lower-inductance modulation loop area, and a flux-coupled, single pickup loop design that replaces previous designs, which were direct-coupled and had two pickup loops.  The final section, section \ref{sec:backaction}, describes methods for reducing the amount of Johnson noise seen by the sample, thus allowing for lower sample temperatures and the possibility of measuring additional mesoscopic effects in small samples.  

\section{Advantages of a planarized device with multi-layer sub-micron features}
\label{sec:planar_design}

The thickness of superconducting layers in an integrated process is constrained to be larger than the superconducting penetration depth.  Layers thinner than this have limited field screening abilities and narrow, thin features can have a non-negligible kinetic inductance.  
To avoid these potential problems, superconducting foundries post design rules with layer thicknesses between 150 and 200 nm, substantially greater than the approximately 85 nm penetration depth of sputtered Niobium.  
Given the thickness of individual layers, it is difficult to design deep sub-micron features into a multi-layered architecture, because small features cannot be reliably fabricated when the underlying height profile varies on the same length scale.  It is therefore reasonable to conclude that any new superconducting foundry that utilizes the now readily available sub-micron lithography tools, will also likely choose to adopt chemical-mechanical polishing steps after each conductor/insulator layer.

\begin{figure}[htp]
\centering
\includegraphics[scale=1]{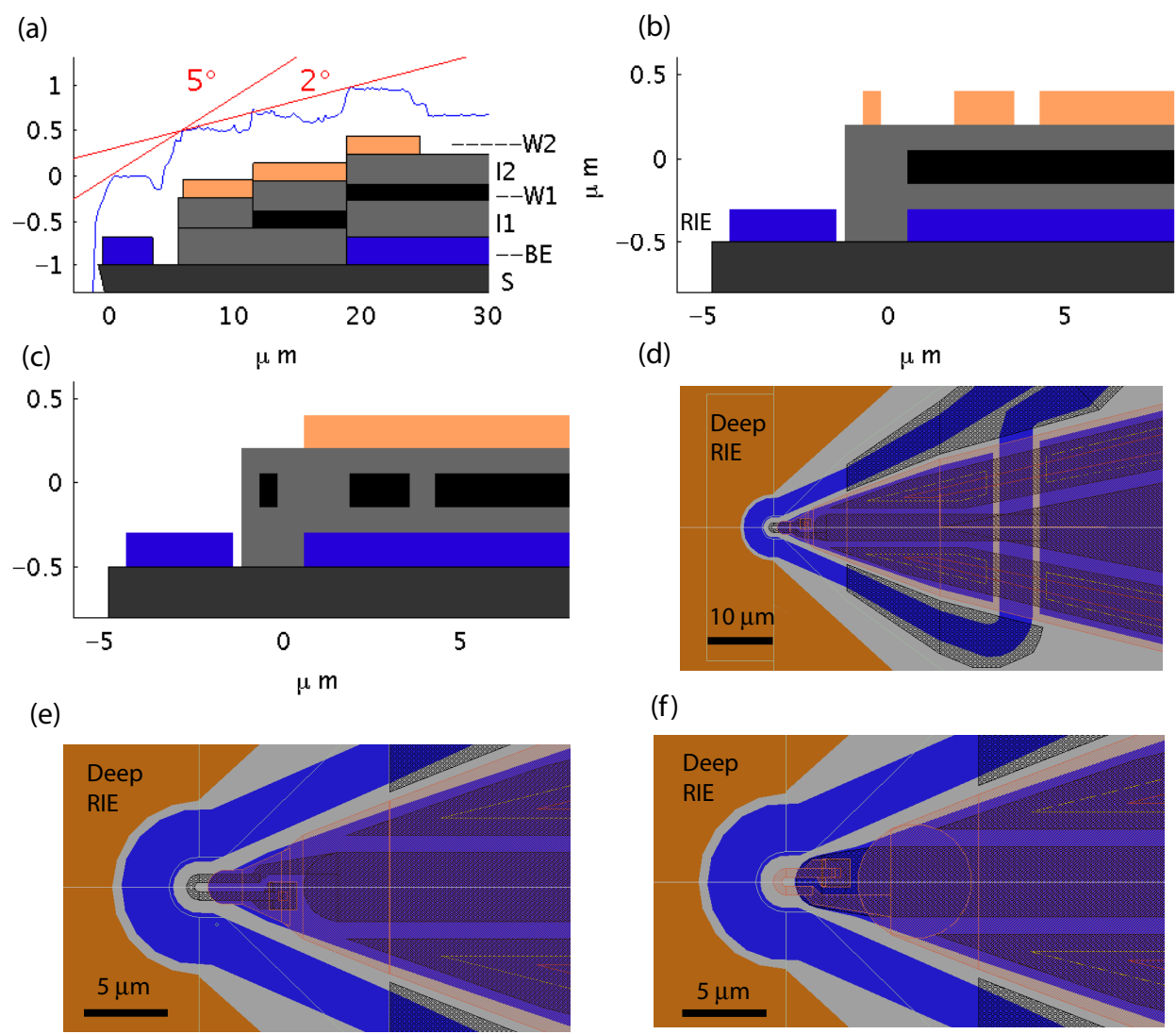}
\caption[Planarized layer effects and an improved sub-micron design]{ (Color Online)
Comparison of design layouts and layer considerations for a planarized vs a non-planarized design with sub-micron feature sizes.  (a-c) Cross-sections down the the center line of the design showing layer thickness effects.  (a) Layer thicknesses of a non planarized process described elsewhere \cite{koshnick_terraced_2008}, with three superconducting layers (BE, W1, W2), two insulating layers (I1, I2), and the substrate.  Atomic force microscope data (blue) shows the tolerance angles that would allow a top layer FIB fabricated pickup loop to touch first.   (b) The layer thicknesses of a planarized design allow for much shorter pickup loop leads with the same tolerance angles.  (c) The layer thickness diagram of a middle layer pickup loop design.   (d-f) Top down view of SQUID design.  (d) Passing the field coil under the linear coax section of the SQUID leads allows for a symmetric tip.  (e) The mid-layer pickup loop design shown in (c).  (f) The top layer design shown in (b).}
\label{fig:ibmSQ_layers}
\end{figure}

The adoption of a planarization process would considerably reduce the geometric constraints that determine whether the pickup loop layer will touch down first.  Our previous paper \cite{koshnick_terraced_2008} described the use of a shallow etch outside the pickup loop and a deep etch outside the field coil.   This geometry, shown in the layer diagram in Fig. \ref{fig:ibmSQ_layers}a, can allow a top-layer pickup loop to touch down first with an alignment tolerance of 3 degrees.   Such alignment tolerance is necessary because of the limited alignment ability in hand-crafted scanners and because changes after alignment due to thermal contractions.   With a planarized process, a minimum alignment angle is no longer inherent in the design.  This means a larger tolerance angle can be achieved (Fig \ref{fig:ibmSQ_layers}b,c,), or a wider field coil can be used, which has the potential of increasing the maximum usable applied field.  

Another very important geometric advantage of a planarized design is that the via to a top layer pickup loop can be much closer to the pickup loop itself, which would allow for a smaller imaging kernel.   In the design shown in Fig \ref{fig:ibmSQ_layers}a, the electrical connection to the top layer is more than 15 $\mu$m from the touch down point.   Although the magnetic field seen by the pickup loop leads (in layer W2) is partially screened by the middle layer (W1), the long leads still make for a large area where the SQUID is responsive to the sample's magnetic field.   In a planarized design, the vias can be brought as close as is allowed by the design rules.  

In previous papers we have demonstrated how susceptometry and background cancellations can enable Johnson noise limited detection of magnetic signals from mesoscopic samples such as superconducting and normal metal rings.   Such objects can have periodic response due to quantum mechanical effects, where the amplitude of this response is inversely related to the ring size.  The minimum ring size is set by the applied field, because at least half a flux quantum of field is required to distinguish a periodic signal from an (often much larger) linear signal related to spins \cite{bluhm_spinlike_2009}.   Furthermore, we have found that if the field is not sourced by the local field coils, then coupling into the SQUID modulation loop, SQUID array amplifier, and additional vibration-related signals severely reduce the technique's sensitivity.   Increasing the width and thickness of the field coil loop is not enough to increase the usable maximum applied field.  We have found that these measures increase the field coil current that makes the SQUID go normal, but that sensitivity is no longer optimal once vortices begin to move at a much lower field \cite{koshnick_terraced_2008}.   Planarization allows for continuous conducting layers, which reduces thin superconducting areas in the design, and thus should help to reduce the field coil current where vortices begin to move.

Spatial resolution is an important factor of a SQUID's design for many reasons.  In the large set of cases where it is acceptable to have the pickup loop $\gtrapprox 300$ nm from the sample, it is advantageous to put the pickup loop in the middle layer as shown in Figs. \ref{fig:ibmSQ_layers}c-e.  This allows for a minimal image kernel size, because both the top and bottom layers can screen the pickup loop from magnetic field. The width of the shielding layers has been kept to a minimum in the area immediately adjacent to the pickup loop to minimize the flux-focusing effects that can increase the effective pickup loop size \cite{ketchenoptimization}.   

Our design for a SQUID with a pickup loop in the top layer, Fig.\ref{fig:ibmSQ_layers}f, is similar to that of Fig.\ref{fig:ibmSQ_layers}e, except that the flux-focusing condition was relaxed to allow for shorter leads and thus less stray coupling.  Both designs have a field coil line that crosses far behind the pickup loop area in a region where the SQUID has a linear coaxial geometry, Fig. \ref{fig:ibmSQ_layers}d.    This allows for a symmetric tip, minimal unshielded leads, and thus a pickup and transition area that contributes minimally to the inductance of the rest of the design.

\section{Optimizing flux sensitivity through reduced inductance}
\label{sec:flux_optimization}

While our previous devices have flux sensitivities that are comparable to or better than scanning devices of their
kind, quantum-limited devices \cite{SQ_handbook} with flux
sensitivities as low as $\sim 0.08 \mu\Phi_0/\sqrt{Hz}$
\cite{awschalom_low-noise_1988} have been realized.  
We note that our designs are limited by thermal
noise in the shunt resistors.  When optimally tuned, this Johnson
noise dependence scales like $L^{3/2}$ where L is the SQUID's self
inductance \cite{tesche_dc_1977}.   An optimally designed quantum limited flux noise
scales as $L^{1/2}$.  
By reducing $L$ one can both reduce the SQUID's overall white noise
floor, and increase the temperature, where temperature-independent
quantum noise becomes the dominate noise source.   

For both the optically \cite{huber_gradiometric_2008} and FIB \cite{koshnick_terraced_2008} fabricated SQUIDs, the inductance \cite{kamon_fasthenry_1994} of each section is approximately 12 pH per pickup loop, 10 pH/mm in the linear coaxial connection region, and $\sim$ 55 pH for the core junction/modulation loop area.  

The majority of the core area inductance in the previous designs come from the two modulation loops, each with an inner diameter of 10 $\mu$m.  Estimating from their inner diameter \cite{jaycox_planar_1981}, these two loops contribute 30 pH to the total inductance.   This contribution would decrease linearly with a reduction in the diameter of the loop.  The main drawback to this change in our case (low bandwidth requirements) is that more modulation-loop current would be required to keep the SQUID in a flux-locked loop.  Increases in the amount of current sent to the modulation loops can lead to heating from stray resistances in dilution refrigerator temperature wiring.  Reducing the modulation loop diameter to 4 $\mu$m decreases the self inductance to 12 pH, while maintaining a reasonable 175 $\mu$A/$\phi_0$ mutual inductance.  Further enhancement could come from having only one modulation loop in a non-gradiometric design.  Small modulation loops with multiple windings could also allow for smaller loops.

Another avenue worth considering for reduced inductance involves questioning the assumption that
dual, counter wound pickup coils aid background subtraction.   Assuming that 
comparative/background measurements are made, either by scanning or by
systematically positioning the sensor at various points around a mesoscopic object, 
the net effect of the counter winding is to reduce the pickup loop coupling to the field
coils to a sufficient extent that the dynamic range of the direct signal does not overwhelm
the room temperature electronics.  For small pickup loops the field coil coupling is already
small, thus the counter-winding may not be necessary.  It is also physically reasonable to 
require room temperature electronics that send some of the field coil signal to the modulation
loop, so that this direct coupled signal is cancelled before amplification. 
It is thus possible that a one-sided SQUID could have lower noise.

\begin{figure}[htp]
\centering
\includegraphics[scale=1]{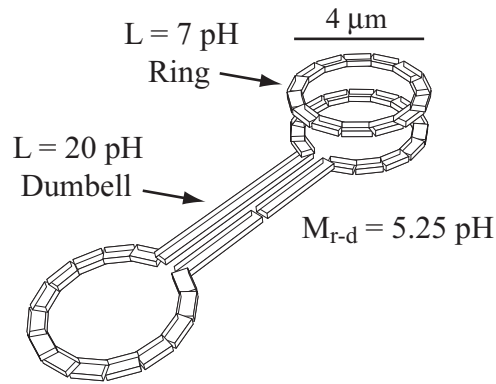}
\caption[Simplified flux coupled SQUID model]{ A simplified model for a flux coupled SQUID, which consists of a dumbell shaped flux coupler with 4 $\mu$m loops, and a 4 $\mu$m ring with specified inductances. }
\label{fig:dumbell}
\end{figure}

Directly cutting out one of the pickup loops creates an unbalanced inductance in the two sides of the SQUID circuit.
This leads to non-symmetric I-V characteristics \cite{peterson_analysis_1979} and slightly more complicated modeling, but does not necessarily reduce the flux sensitivity.  One way to reduce the unbalanced inductance problem would be to remove the modulation loops all together.  In the resulting two-pickup loop structure, the non-scanning pickup loop could be associated with an additional field coil that acts as a modulation loop.  
 
An alternative design direction is to change the bulk of the current design into a flux coupler with 
just the central part being an independent low-inductance SQUID. 
A simplified Fasthenry  \cite{kamon_fasthenry_1994} model of superconducting elements (Fig \ref{fig:dumbell}) illustrates the tradeoffs between the reduced coupling to the sample, and the reduction in the SQUID's inductance. 
It considers the inductances of two objects: a simple 4 $\mu$m diameter ring, and a ``dumbell" consisting of two 4 $\mu$m diameter loops coupled by two lines.   
This dumbell object has a self inductance, $L_d =20$ pH, and 
can be considered either as a direct coupled SQUID or the flux coupler that is coupled to
the ring.  In the latter case, the ring, which has a self inductance $L_r = 7$ pH, represents the
non-directly coupled SQUID.   
The mutual inductance between the ring and dumbell is $M_{r-d} = 5.25$ pH.
The flux-coupled SQUID only sees $M_{r-d}/L_d =.26$ as much flux as its direct-coupled
counterpart, but when limited by Johnson noise, it has $(L_r/L_d)^{3/2} = .21$ as much white noise.
Flux-coupled and direct-coupled SQUIDs therefore have very similar white noise characteristics
until the inductance is small enough that the $L^{1/2}$-dependent quantum limit is applicable.
The main advantage may be that a flux-coupled SQUID could have only one pickup loop, 
substantially reducing the inductance without leaving the SQUID unbalanced.

\section{Reducing SQUID sample back action}
\label{sec:backaction}

Understanding the way the SQUID perturbs the sample is an important part of 
measuring quantum effects in mesoscopic samples.   We will consider one of the simplest kinds of 
back action on the sample, joule heating from SQUID radiation at the Josephson frequency, $f_j$.  
While studying persistent currents in normal metal rings, we discovered that metallic
regions had a spin-like 
linear susceptibility that increased with decreasing temperature \cite{bluhm_spinlike_2009}.  The coupled 
SQUID radiation of $ \approx 10^{-14}$ Watts for our $R \approx 2 \Omega$ rings, balanced by electron-phonon limited cooling \cite{wellstood_hot-electron_1994}, limited the 
temperature in isolated rings to about 200 mK \cite{bluhm_persistent_2009}.
Since there are many interesting effects that only occur below this temperature, 
it is important to consider design aspects that can reduce this form of back action.

In general, the cost of reduced back-action is often reduced sensitivity.   
A simple estimate for the energy dissipated in a ring is given by the flux-change-induced
ring-voltage
\begin{equation}
V_{ring} = \frac{d\Phi}{dt} = I_0 f_j M_{SQ-s}
\end{equation}
where $M_{SQ-s}$ is the mutual inductance between the SQUID and the sample.  
The power dissipated is then just $V_{ring}^2/R$, where R is the ring's resistance.
One can reduce this heating effect by moving the SQUID further away
from the sample, thereby reducing $M_{SQ-s}$.   This is only a useful approach
when there is ample signal.   Another approach is to bias the SQUID with the lowest possible
voltage, and thus lowest operating $f_j$.   In general, this approach is limited by 
the minimum voltage bias ($\propto k_BT/e$), where the SQUID noise begins to deteriorate, and by
the feedback electronics' ability to lock into a signal that has a reduced modulation amplitude.

The noise dependance on $L^{3/2}$, mentioned in section \ref{sec:flux_optimization}, 
was calculated in the
$\beta_L = 2 L I_0 / \Phi_0 \approx 1$ limit, where the standard design rules  \cite{tesche_dc_1977}
set critical current  to $I_0 = \Phi_0/2L$, and the appropriate shunt
resistor is determined by $\beta_c \approx 1$.  Increasing the flux sensitivity
by reducing $L$ thus comes with an increased $I_0$ and correspondingly increased radiation 
on the sample.   While the SQUID's energy sensitivity, $\Phi_n^2/L$, is optimal
for $\beta_L \approx 1$, the relevant figure of merit when measuring
mesoscopic samples must include back action.   The optimal point 
in this case would probably require $I_0 < \Phi_0 / 2 L$.   
This effect has been analyzed in quantum limited SQUIDs \cite{danilov_quantum_1983}.

Another way to reduce the back action on the sample is to read out the
SQUID dispersively.   Josephson radiation (and thus sample heating) does 
not occur when the SQUID is in the zero voltage state.   Dispersive readout measures
the flux-dependence of the inductance, without inducing voltage, typically by placing the 
SQUID in some kind of resonant circuit.   Thus the Josephson radiation (of typically 1-10 GHz) gets replaced by the readout frequency ($\sim$ 100 Mhz).  One excellent
way to implement dispersive readout involves the microstrip SQUID amplifier
designed by the group of John Clarke \cite{muck_superconducting_2001}.  
The superconducting qubit community has invested considerable effort in
comparing the back action from this type of measurement to the back action
from other non-voltage state readout schemes, such as the scanning SQUIDs \cite{hasselbach_microsquid_2000} that rely on the bias current switching threshold.  

A final way to limit SQUID heating would be to implement on-chip filters 
inside the SQUID circuit that divert the Josephson frequency currents from the
pickup loop and sample.  This approach was 
implemented with LC filters by John Price's  group \cite{Zhang} in a two-chip design.
Implementing the same type of LC filters on a single chip would help to keep 
inductive losses to a minimum.   We note
the capacitance in our 40 $\mu$m wide, 1.2 mm long linear coaxial coupler 
with a 200 nm wire-to-wire spacing is roughly 20 pF.  The 30  pH pickup loop and 
strip line inductance thus has a cutoff frequency of 6.5 Ghz, indicating that a 
reasonable portion of the Josephson frequency current is already not flowing through
the pickup loop itself.   We therefore believe that further design work that explicitly exploits 
this effect could have significant success. 

A second type of on-chip filter involves placing a placing a small (~0.3 $\Omega$) shunting resistor between the SQUID and the pickup loop.  This form of RC filter works as a frequency dependent current divider, where the Josephson frequency currents would see a reactance of 0.95 $\Omega$ (15 pH at $\sim$ 10 Ghz) and thus largely run through the shunt, whereas the zero frequency currents that set up the SQUID phase relations would still all be coupled to the pickup loop itself.  

As with the LC filter technique, this line of reasoning intrinsically relies on the fact that we are trying to measure time-averaged equilibrium currents in our rings, and thus can simply limit the frequencies where effective coupling occurs to some value below the frequencies involved with the intrinsic SQUID and sample dynamics.  The LC filter could effectively shunt the great majority of the Josephson frequency radiation if an additional high capacitance layer, with sufficiently low stray inductance, was incorporated into the superconducting process.  The effectiveness of the RC filter is limited by the Johnson current noise it sends through the pickup loop, integrated up to the bandwidth set by the LC filter.  In practice, however, it is likely that this RC shunting technique could reduce the sample temperature significantly before this limit sets in.

This work was supported by NSF Grants Nos. DMR-0803974 and PHY-0425897 

\section*{References}


\begin{thebibliography}{10}

\bibitem{ketchenoptimization}
M.~B. Ketchen and J.~R. Kirtley.
\newblock Design and performance aspects of pickup loop structures for
  miniature squid magnetometry.
\newblock {\em IEEE Transactions on Applied Superconductivity}, 5(2):2133 --
  2136, Jun 1995.

\bibitem{kirtley_high-resolution_1995}
J.~R. Kirtley, M.~B. Ketchen, K.~G. Stawiasz, J.~Z. Sun, W.~J. Gallagher, S.~H.
  Blanton, and S.~J. Wind.
\newblock High-resolution scanning squid microscope.
\newblock {\em Applied Physics Letters}, 66:1138--1140, February 1995.

\bibitem{tsuei_pairing_1994}
C.~C. Tsuei, J.~R. Kirtley, C.~C. Chi, L.~S. {Yu-Jahnes}, A.~Gupta, T.~Shaw,
  J.~Z. Sun, and M.~B. Ketchen.
\newblock Pairing symmetry and flux quantization in a tricrystal
  superconducting ring of {YBa}$_2${Cu}$_3${O}$_{7-delta}$.
\newblock {\em Physical review letters}, 73(4):593--596, 1994.

\bibitem{kirtley2006}
J.R. Kirtley, C.C. Tsuei, Ariando, C.J.M. Verwijs, S.~Harkema, and
  H.~Hilgenkamp.
\newblock Angle-resolved phase-sensitive determination of the in-plane gap
  symmetry in {YBa}$_2${Cu}$_3${O}$_{7- \delta}$.
\newblock {\em Nature Physics}, 2:190, 2006.

\bibitem{moler_images_1998}
Kathryn~A. Moler, John~R. Kirtley, D.~G. Hinks, T.~W. Li, and Ming Xu.
\newblock Images of interlayer josephson vortices in
  {Tl}$_2${Ba}$_2${CuO}$_{6+\delta}$.
\newblock {\em Science}, 279(5354):1193--1196, February 1998.

\bibitem{bluhm_persistent_2009}
Hendrik Bluhm, Nicholas~C. Koshnick, Julie~A. Bert, Martin~E. Huber, and
  Kathryn~A. Moler.
\newblock Persistent currents in normal metal rings.
\newblock {\em Physical Review Letters}, 102(13):136802--4, April 2009.

\bibitem{gardner_scanning_2001}
Brian~W. Gardner, Janice~C. Wynn, Per~G. Bjornsson, Eric W.~J. Straver,
  Kathryn~A. Moler, John~R. Kirtley, and Mark~B. Ketchen.
\newblock Scanning superconducting quantum interference device susceptometry.
\newblock {\em Review of Scientific Instruments}, 72(5):2361--2364, 2001.

\bibitem{koshnick_fluctuation_2007}
Nicholas~C. Koshnick, Hendrik Bluhm, Martin~E. Huber, and Kathryn~A. Moler.
\newblock Fluctuation superconductivity in mesoscopic aluminum rings.
\newblock {\em Science}, 318:1440--1443, November 2007.

\bibitem{hilgenkamp_ordering_2003}
Hans Hilgenkamp, Ariando, {Henk-Jan}~H. Smilde, Dave H.~A. Blank, Guus
  Rijnders, Horst Rogalla, John~R. Kirtley, and Chang~C. Tsuei.
\newblock Ordering and manipulation of the magnetic moments in large-scale
  superconducting $\pi$-loop arrays.
\newblock {\em Nature}, 422(6927):50--53, March 2003.

\bibitem{tafuri_magnetic_2004}
F~Tafuri, J~R Kirtley, P~G Medaglia, P~Orgiani, and G~Balestrino.
\newblock Magnetic imaging of pearl vortices in artificially layered
  ({Ba}$_0.9${Nd}$_0.1${CuO}$_{2+x}$ )$_m$/({CaCuO}$_2$)$_n$ systems.
\newblock {\em Physical Review Letters}, 92(15):157006, April 2004.
\newblock {PMID:} 15169312.

\bibitem{huber_gradiometric_2008}
Martin~E. Huber, Nicholas~C. Koshnick, Hendrik Bluhm, Leonard~J. Archuleta,
  Tommy Azua, Per~G. Bj\"{o}rnsson, Brian~W. Gardner, Sean~T. Halloran, Erik~A.
  Lucero, and Kathryn~A. Moler.
\newblock Gradiometric micro-squid susceptometer for scanning measurements of
  mesoscopic samples.
\newblock {\em Review of Scientific Instruments}, 79(5):053704, 2008.

\bibitem{koshnick_terraced_2008}
Nicholas~C. Koshnick, Martin~E. Huber, Julie~A. Bert, Clifford~W. Hicks, Jeff
  Large, Hal Edwards, and Kathryn~A. Moler.
\newblock A terraced scanning superconducting quantum interference device
  susceptometer with submicron pickup loops.
\newblock {\em Applied Physics Letters}, 93(24):243101--3, December 2008.

\bibitem{ketchen_sub-mu_1991}
M.~B. Ketchen, D.~Pearson, A.~W. Kleinsasser, {C.-K.} Hu, M.~Smyth, J.~Logan,
  K.~Stawiasz, E.~Baran, M.~Jaso, T.~Ross, K.~Petrillo, M.~Manny, S.~Basavaiah,
  S.~Brodsky, S.~B. Kaplan, W.~J. Gallagher, and M.~Bhushan.
\newblock Sub-mu m, planarized, {Nb-AlO}$_x${-Nb} josephson process for 125 mm
  wafers developed in partnership with {Si} technology.
\newblock {\em Applied Physics Letters}, 59(20):2609--2611, November 1991.

\bibitem{bluhm_spinlike_2009}
Hendrik Bluhm, Julie~A. Bert, Nicholas~C. Koshnick, Martin~E. Huber, and
  Kathryn~A. Moler.
\newblock Spinlike susceptibility of metallic and insulating thin films at low
  temperature.
\newblock {\em Physical Review Letters}, 103(2):026805, 2009.

\bibitem{SQ_handbook}
J.~Clarke and A.~I. Braginski, editors.
\newblock {\em The SQUID Handbook}.
\newblock Weinheim : Wiley-VCH, 2003.

\bibitem{awschalom_low-noise_1988}
D.~D. Awschalom, J.~R. Rozen, M.~B. Ketchen, W.~J. Gallagher, A.~W.
  Kleinsasser, R.~L. Sandstrom, and B.~Bumble.
\newblock Low-noise modular microsusceptometer using nearly quantum limited dc
  squids.
\newblock {\em Applied Physics Letters}, 53(21):2108--2110, November 1988.

\bibitem{tesche_dc_1977}
Claudia~D. Tesche and John Clarke.
\newblock dc squid: Noise and optimization.
\newblock {\em Journal of Low Temperature Physics}, 29:301--331, November 1977.

\bibitem{kamon_fasthenry_1994}
M.~Kamon, M.~J. Tsuk, and J.~K. White.
\newblock Fasthenry: a multipole-accelerated 3-d inductance extraction program.
\newblock {\em IEEE Transactions on Microwave Theory and Techniques},
  42(9):1750 -- 1758, Sept 1994.

\bibitem{jaycox_planar_1981}
J.~Jaycox and M.~Ketchen.
\newblock Planar coupling scheme for ultra low noise {DC} {SQUIDs}.
\newblock {\em Magnetics, {IEEE} Transactions on}, 17(1):400--403, 1981.

\bibitem{peterson_analysis_1979}
R.~L. Peterson and C.~A. Hamilton.
\newblock Analysis of threshold curves for superconducting interferometers.
\newblock {\em Journal of Applied Physics}, 50:8135--8142, December 1979.

\bibitem{wellstood_hot-electron_1994}
F.~C. Wellstood, C.~Urbina, and John Clarke.
\newblock Hot-electron effects in metals.
\newblock {\em Physical Review B}, 49:5942, March 1994.

\bibitem{danilov_quantum_1983}
V.~Danilov, K.~Likharev, and A.~Zorin.
\newblock Quantum noise in squids.
\newblock {\em Magnetics, IEEE Transactions on}, 19(3):572--575, 1983.

\bibitem{muck_superconducting_2001}
Michael Muck, J.~B. Kycia, and John Clarke.
\newblock Superconducting quantum interference device as a near-quantum-limited
  amplifier at 0.5 ghz.
\newblock {\em Applied Physics Letters}, 78(7):967--969, February 2001.

\bibitem{hasselbach_microsquid_2000}
K.~Hasselbach, C.~Veauvy, and D.~Mailly.
\newblock Microsquid magnetometry and magnetic imaging.
\newblock {\em Physica C}, 332:140--147, May 2000.

\bibitem{Zhang}
X.~Zhang and J.~C. Price.
\newblock Susceptibility of a mesoscopic superconducting ring.
\newblock {\em Phys.\ Rev.\ B}, 55(5):3128 -- 3140, FEB 1997.

\end{thebibliography}

\end{document}